\documentclass[letterpaper,aps,prl,twocolumn]{revtex4-1}

\usepackage{graphicx}
\usepackage{amssymb,amsmath}
\usepackage{braket}
\begin{document}

\newcommand{\Buzeketal}{Bu\ifmmode \check{z}\else \v{z}\fi{}ek \textit{et al.}}

\title{Exponential scaling of clock stability with atom number}
\author{T. Rosenband}
\email[]{trosen@boulder.nist.gov}
\author{D. R. Leibrandt}
\affiliation{National Institute of Standards and Technology, 325
Broadway, Boulder, CO 80305}

\date{\today}

\begin{abstract}
In trapped-atom clocks, the primary source of decoherence is often the phase noise of the oscillator.  For this case, we derive theoretical performance gains by combining several atomic ensembles.  For example, $M$ ensembles of $N$ atoms can be combined with a variety of probe periods, to reduce the frequency variance to $M 2^{-M}$ times that of standard Ramsey clocks.  A similar exponential improvement is possible if the atomic phases of some of the ensembles evolve at reduced frequencies.  These ensembles may be constructed from atoms or molecules with lower-frequency transitions, or generated by dynamical decoupling.  The ensembles with reduced frequency or probe period are responsible only for counting the integer number of $2\pi$ phase wraps, and do not affect the clock's systematic errors.  Quantum phase measurement with Gaussian initial states allows for smaller ensemble sizes than Ramsey spectroscopy.
\end{abstract}
\maketitle

Atomic clocks have many technical and scientific applications, and also serve as model systems for quantum metrology.  An important result of quantum mechanics is that the variance of phase measurements can be reduced if entangled particles are used rather than an equal number of unentangled ones~\cite{Caves1981,Giovannetti2004}.  This result has been applied to the theoretical analysis of atomic clocks that operate by repeatedly measuring the phase difference between atomic qubits and a classical oscillator~\cite{WMI1993ProjectionNoise,Huelga1997,DJW1998bible,Buzek1999Optimal,riis2004optimum,Lukin2004Stability,Rosenband2011NumTest,Mullan2012Semidefinite}.  However, these analyses have not taken full advantage of the hybrid nature of passive atomic clocks, where a classical oscillator with limited coherence reproduces the resonance frequency of more coherent quantum systems.  The present model derives an exponential gain from additional couplings between atoms and the laboratory system, to digitize their relative phase evolution.  It is related to classical analog-to-digital conversion techniques~\cite{kester2005data} and digital computation of the $arctan$ function~\cite{Volder1959CORDIC}, which have been extended to the use of qubits for the measurement of classical fields~\cite{Vaidman2004} and mesoscopic spins~\cite{Giedke2006}.  Here it is shown that if the information from $M$ ensembles of $N$ atoms is combined classically, the frequency variance may be reduced to $M D^{1-M}/2$ times the variance of standard Ramsey clocks with $MN$ atoms, where the value of $D$ is constrained by the ensemble size.  The method works by extending the range of invertible phases from the usual $-\pi\ldots\pi$ interval to $-D^{M-1}\pi\ldots D^{M-1}\pi$. One ensemble is used for normal phase measurements, while the other $M-1$ ensembles are measured with free-evolution periods that are reduced by factors $D^{-1},\ldots,D^{1-M}$, or evolve at frequencies that are reduced by the same factors.  This allows unambiguous counting of the number of $2\pi$ phase wraps that occur in the first ensemble.

The passive atomic clocks considered here operate by periodically measuring the differential phase evolution between a classical oscillator and an ensemble of atoms.  These measurements yield corrections to the oscillator, which is adjusted in a feedback loop to produce an output signal whose frequency approximates that of the ideal unperturbed atomic resonance.  The stability of atomic clocks, which describes the uncertainty of this approximation as a function of the averaging duration, is limited by quantum projection noise~\cite{WMI1993ProjectionNoise} for short durations.  If a clock operates by repeatedly applying the Ramsey technique~\cite{Ramsey1956} to measure the phase difference of $N$ unentangled atoms with respect to the oscillator, the projection-noise-limited fractional-frequency stability is $\sigma_{yp}(\tau)=(\omega\sqrt{N T \tau})^{-1}$, where $T$ is the free-evolution period, $\tau$ is the total duration, and $\omega$ is the angular frequency of the atoms~\cite{WMI1993ProjectionNoise,Huelga1997}.  This stability limit is an important figure of merit, and it connects atomic clocks to quantum metrology, because the use of entangled states allows $\sigma_{yp}(\tau)$ to scale more favorably as $N^{-1}$.  A clock that operates for duration $\tau$ has a fractional time-variance of  $var(\tau)/\tau^2=\sigma_{yp}(\tau)^2+\sigma_{ya}(\tau)^2$, where $\sigma_{ya}(\tau)$ is the accuracy, which does not decrease below the bounds set by the systematic errors.  For short measurement durations, quantum measurement noise restricts the degree to which atomic clocks can reproduce the ideal frequency $\omega$.  For long durations, the dominant uncertainty comes from systematic errors due to effects such as unknown environmental factors.

Proposals to use the available atoms more efficiently, and thereby reduce $\sigma_{yp}$, include the use of spin-squeezed states~\cite{Kitagawa1993Squeezed, Wineland1994Squeezed, Lukin2004Stability, Vuletic2010Squeeze}; more general quantum measurements are also possible~\cite{JJB1996Entanglement,Buzek1999Optimal,Dobrzanski2011,Rosenband2011NumTest,Mullan2012Semidefinite}.  Such approaches are restricted by the fact that the atomic phase evolves as $\phi=\int^T_0 (\omega-\Omega(t))dt$ with respect to the classical oscillator, whose frequency is $\Omega(t)$.  When the atoms evolve freely, there is no way to unambiguously resolve phases beyond $\pm\pi$.  Because of this, $T$ is usually constrained to a duration that is short enough so that the atom-oscillator phase difference has not drifted outside the $-\pi$ to $\pi$ range of invertibility.  Here it is shown that the range of phase invertibility and therefore $T$ can grow exponentially when multiple atomic ensembles are available.

Consider a passive atomic clock as described above.  Atomic decoherence is assumed to be negligible compared to oscillator decoherence.  This idealization approximates many current atomic clocks, where the oscillator is a laser whose output has a $1/f$ power spectrum of frequency fluctuations~\cite{Numata2004ThermalNoise}.  Then the phase difference $\phi$ can be thought of as a Gaussian random variable with standard-deviation $\sigma_\phi=\alpha T$, where $\alpha$ characterizes the oscillator noise level.  Only phases inside some interval $-\theta\ldots\theta$ can be unambiguously inverted, and it is reasonable to require that $\sigma_\phi\le \theta/6$, making phase-wrapping errors six-sigma events.  Other choices of the ratio $\sigma_\phi/\theta$ are also possible, to reduce this probability to a level that is considered insignificant.  In this Letter, such a clock is called a standard Ramsey clock.  Here the limit is $\theta=\pi/2$ and $T = \pi / {12 \alpha}$, because the clock measures a single quadrature of $\phi$.  For $MN$ atoms the projection-noise-limited fractional frequency stability is
$\sigma_{yp}(\tau)=\sqrt{12\alpha}/(\omega\sqrt{MN \pi \tau})$.  

It has been suggested~\cite{Buzek1999Optimal, Lloyd2013Private} that if $M$ oscillators with frequencies $\omega$, $\omega/2$,\ldots,$\omega/2^{M-1}$ are available, then durations up to $2^{M}\pi/\omega$ can be measured with the precision of the fastest oscillator.  Here we show that when $M$ atomic ensembles are probed with reduced free-evolution periods or which evolve at reduced frequencies, the range of unambiguous phase excursions is extended to $\theta=D^{M-1}\pi$, where $D^0,D^1,\ldots,D^{M-1}$ are either probe-time or frequency-division ratios.  Then the Ramsey free-evolution period in the presence of $1/f$ oscillator noise can grow to $T = D^{M-1}\pi/(6\alpha)$. The projection-noise limited stability for the the $N$ unperturbed atoms is then $\sigma_{yp}(\tau)=\sqrt{6\alpha}/(\omega\sqrt{D^{M-1} N \pi \tau})$.  This represents a reduction in the clock's frequency variance to $MD^{1-M}/2$ times the variance of a standard Ramsey clock with the same number of atoms.

The hypothetical clock contains $M$ ensembles of $N$ atoms.  Atoms are thought of as two-level systems or qubits.  Each ensemble $A_j$ is divided into two sub-ensembles of $N/2$ atoms, labeled $X_j$ and $Y_j$, where $j\in{0...M-1}$.  The atoms of all $X_j$ are simultaneously prepared in the quantum ground state, and probed by a Ramsey sequence that consists of a $\pi/2$-pulse about the Bloch-sphere x-axis, followed by evolution for a period $T$.  During this period, the atoms evolve through a reduced phase $\phi_j=\int^T_0 (\omega-\Omega(t))/D^j dt$ with respect to the oscillator.  A possible construction by use of dynamical decoupling is described later. The evolution period ends with a second $\pi/2$-pulse about the x-axis, and final measurement along the z-axis.  The measurement result is recorded as the fraction of atoms $x_j$ in the excited state.  The $Y_j$ sub-ensembles are simultaneously probed in the same way, except that the second $\pi/2$-pulse is about the y-axis, and their results are recorded as $y_j$.  The expectation values $\langle x_j \rangle=(1+\cos{\phi_j})/2$ and $\langle y_j \rangle=(1+\sin{\phi_j})/2$ are quadratures of $\phi_j$, and the value $\beta_j=arg((x_j-1/2)+i (y_j-1/2))$ is a 
measurement of $\phi_j$, modulo $2\pi$.  Here $arg$ is the argument function, whose range is $-\pi\ldots\pi$.

First consider the phase uncertainty of each ensemble.  The measurements $x_j$ and $y_j$ yield an estimate of the phase $\phi_j$.  This can be written as $\phi_j=\beta_j+2P_j\pi+G_j$, where $P_j$ is an integer whose value may be unknown, and $G_j$ is a random variable that completes the equation.   The distribution of $G_j$ values is determined by the $arg$ function and the scaled binomial random variables $x_j$ and $y_j$.  In the limit of large $N$, $G_j$ approximates a Gaussian random variable with variance $\sigma^2=1/N$, independently of the actual phases~\cite{WMI1993ProjectionNoise}.  To determine the ensemble size $N$ that is required to reduce the probability of inversion errors to a certain level, we perform precise numerical calculations and do not use the Gaussian approximation.                                                                                                                     

\begin{figure*}
\includegraphics[width=7in]{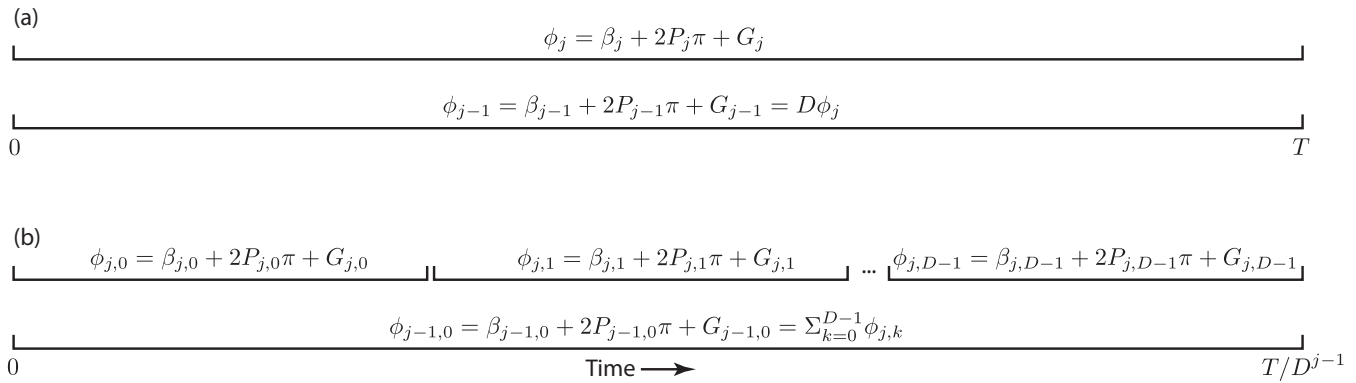}
\caption{Depiction of how phase measurements can be combined for improved clock stability.  In (a) two ensembles with frequency ratio $D$ evolve synchronously for period $T$, and their phase measurements are then combined (see Eq.~1) to derive the integer number of $2\pi$ phase wraps $P_{j-1}$ of the second ensemble.  In (b) the first ensemble is measured $D$ times during each measurement of the second ensemble.  This information also yields the number of phase wraps.  Here the second index is an integer with range $0\ldots (D^j-1)$ and refers to the sequence of measurements during one period $T$.}
\label{FIGSCH}
\end{figure*}

\begin{figure}
\includegraphics[width=3.4in]{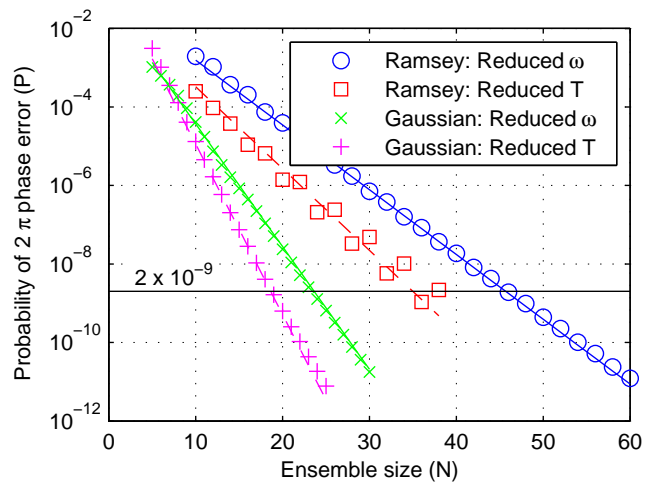}
\caption{Probability of $2\pi$ phase errors $P$ as a function of ensemble size $N$.  Circles show the total probability that the term $|(2G_{j}-G_{j-1})/(2\pi)|$ in Eq.~1 exceeds $1/2$, based on the binomial distributions of $x_j, y_j, x_{j-1}$ and $y_{j-1}$, averaged uniformly over the interval $-\pi\leq\phi_j\leq\pi$.  The calculated probabilities are approximated by the fit $P=0.07 e^{-0.38 N}$.  Squares represent the probability that $2\pi$ phase errors occur when reduced probe periods are used.  In this case the fit is $P=0.04 e^{-0.48 N}$.  The error probabilities for Gaussian initial states measured in the phase-state basis are shown as ($\times$) and (+), and their fits are $P=0.05 e^{-0.72 N}$ and $P=0.25 e^{-N}$ respectively.}
\label{FIGPERF}
\end{figure}

Results from measuring the $M$ ensembles are combined to yield an estimate of the total phase $\phi_0$.  This process can be understood iteratively from lowest frequency to highest, as shown in Fig.~~\ref{FIGSCH}.  Consider the ensembles $A_j$ and $A_{j-1}$ where the integer $P_j$ is assumed to be known, and $P_{j-1}$ is unknown, but can be derived from the earlier equations involving $\phi_j$ as:
\begin{equation}
P_{j-1} = DP_j+\frac{D\beta_j-\beta_{j-1}}{2\pi}+\frac{D G_j - G_{j-1}}{2\pi}
\end{equation}
In practice, $P_{j-1}$ will be calculated as $r(DP_j+\frac{D\beta_{j}-\beta_{j-1}}{2\pi})$ where $r()$ rounds its argument to the nearest integer.  The probability of assigning an incorrect value to $P_{j-1}$ is equal to the probability that $|DG_{j}- G_{j-1}|\geq\pi$, which can be made arbitrarily small by choice of ensemble size $N$ (see Fig.~\ref{FIGPERF}).  Therefore, if the integer $P_j$ is known, then the integer $P_{j-1}$ can be derived from the measurements.  By induction, it follows that the number of phase wraps $P_0$ is known, because the evolution time was chosen such that $P_{M-1}$ is known to be zero.  If errors in the determination of $P_{j-1}$ from $P_{j}$ are equivalent to a six-sigma Gaussian events ($2\times 10^{-9}$ probability), then $N=46$ is the minimum ensemble size when $D=2$ (see Fig.~\ref{FIGPERF}).  The total error probability during each period $T$ is equal to $M-1$ (the number of applications of Eq.~1) times the probability shown in Fig~\ref{FIGPERF}.

Reduced-frequency clocks might be constructed from atoms or molecules with lower-frequency resonances, where the oscillator is derived by frequency division of the original oscillator~\cite{SAD2000Comb, Schnatz1996CohLink}.  In this case, the frequency ratios between ensembles $\phi_{j-1}/\phi_j$ will be given by the natural frequency ratios of the transitions, rather than the equal ratios $D$ considered so far, but the previous arguments still apply, and we can derive a clock variance that is reduced by the product of the transition-frequency ratios.  

A sequence of $\pi$-pulses, known as dynamical-decoupling, can also reduce the rate of phase evolution.  Optimized pulse sequences have been developed to extend the coherence time of qubits, with the goal of making their phase evolution with respect to the oscillator as close to zero as possible.  The same approach that rephases qubits for frequency drifts of polynomial order $n$, by $n+1$ $\pi$-pulses~\cite{Uhrig2007}, can reduce the phase evolution to $\phi_j=\phi_0D^{-j}$.  For example, the reduced phases can be generated in the face of linear drift of the oscillator frequency $\Omega(t)=\Omega_0+t\Omega_1$ during a period $T_D$ by one $\pi$-pulse at $T_A=(T_D/4)(1+D^{-j})$, and a second $\pi$-pulse at $T_B=(T_D/4)(3-D^{-j})$.  In this case, $\phi_j=\int_0^{T_A}(\omega-\Omega(t))dt-\int_{T_A}^{T_B}(\omega-\Omega(t))dt+\int_{T_B}^{T_D}(\omega-\Omega(t))dt=\int_0^{T_D}(\omega-\Omega(t))D^{-j}dt$, as is required.  Higher-order sequences can be formed by adapting the recipe of ref.~\cite{Uhrig2007}. The total evolution period $T$ can contain many decoupling sequences of length $T_D$ to closely approximate the ideal value of $\phi_j$.  In real clocks, generating dynamical decoupling sequences with sufficient fidelity is likely to be difficult.

\begin{figure}
\includegraphics[width=3.4in]{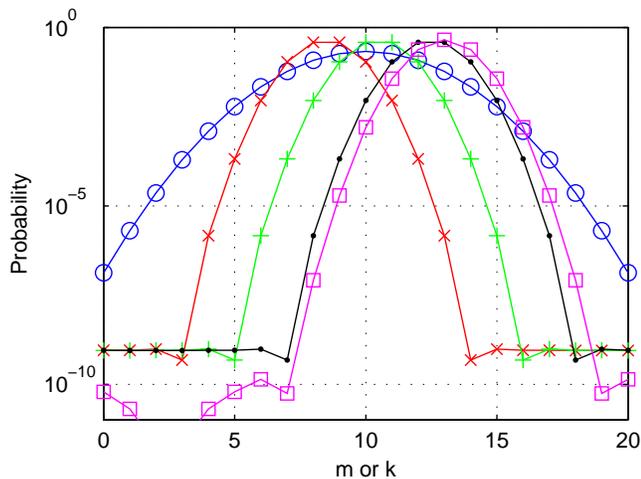}
\caption{Probabilities of $\ket{N,m}$ and $\ket{k}$ for phase measurement with Gaussian states. Lines are drawn to guide the eye.  The initial state $\ket{a}$ for $N=20$ and $c=0.7$ in the $\ket{N,m}$ basis is represented by (o).  Symbols $\times$ + . $\square$ show the probabilities of measuring $\ket{k}$ after phase evolution by $-2\delta, 0, 2\delta,$ and $(5/2)\delta$ respectively, where $\delta=2\pi/(N+1)$.  Phase shifts correspond to translation of $k$.  Large phase measurement errors are confined to the low-probability tails of the Gaussian function.  For the discrete Fourier transform, the product of Gaussian widths in $m$ and $k$ space is proportional to $N$.  The choice of $c$ is a compromise because the width in $k$ space should be as narrow as possible to avoid large phase errors, but cutting off the Gaussian in $m$ space adds unwanted Fourier components.}
\label{FIGGAUSS}
\end{figure}

As an alternative to the quadrature Ramsey measurements discussed above, one can consider more general quantum measurements on each ensemble to derive the phase estimates $\beta_j$.  Optimal measurements will minimize the probability of large phase errors.  First prepare an initial state $\ket{a}$ and after evolution through a phase $\phi_j$, measure in some basis.  \Buzeketal~\cite{Buzek1999Optimal} suggest a measurement basis spanned by the states~\cite{peres1980measurement} $\ket{k}=\frac{1}{\sqrt{N+1}}\sum_{m=0}^{N}e^{i m 2\pi k/(N+1)}\ket{N,m}$ where $\ket{N,m}$ are the fully symmetrized states with $N$ qubits and $m$ excitation (e.g. $\ket{3,2}=(\ket{011} + \ket{101} + \ket{110})/\sqrt{3}$).  The probability amplitude of measuring $\ket{k}$ is simply the $k$th coefficient of the discrete Fourier transform of the values $\bra{a}e^{-im\phi_j}\ket{N,m}$, where $m\in 0\ldots N$ (see Fig.~\ref{FIGGAUSS}).  For the Gaussian initial states~\cite{Lukin2004Stability} where $\ket{a}=(-1)^m e^{-(m-N/2)^2/(N c)}\ket{N,m}$ (unnormalized) the error rate is reduced compared to Ramsey measurements with an equal number of qubits (see Fig.~\ref{FIGPERF}).  Numerically, we find an optimal value of $c=0.735$ and in this case $N=24$ is the smallest number of atoms that yields an error rate below $2\times10^{-9}$ when reduced frequency ensembles with $D=2$ are combined.  While they are not necessarily optimal, the Gaussian states were chosen because the tail of a Gaussian function rapidly approaches zero, for both the original function and its discrete Fourier transform.  This reduces the probability of large phase errors.

Exponential scaling behavior is also possible if the atoms of all $M$ ensembles have frequency $\omega$, but each ensemble $A_j$ freely evolves for a reduced period $T_j = D^{-j} T$ and is then measured and reprepared.  In this case, $D$ is an integer and measurement and preparation are assumed to be instantaneous.  Here $D$ measurements of duration $T_j$ from ensemble $A_j$ are combined to derive the integer number of phase wraps that occur during a single measurement of ensemble $A_{j-1}$ (see Fig.~\ref{FIGSCH}).  Measurements are iteratively combined as before.  Numerical evaluation shows that if $D=2$ and $2\pi$ phase errors are again allowed to occur with $2\times10^{-9}$ probability, then $N=36$ and $N=19$ are the minimum ensemble sizes for Ramsey and Gaussian ($c=0.635$) measurements respectively (see Fig.~\ref{FIGPERF}).  In this case, the total error probability during each period $T$ is equal to $2^{M-1}-1$ times the probability shown in Fig~\ref{FIGPERF}.

The above techniques show that oscillator phase noise in atomic clocks may be mitigated by use of a collection of atomic ensembles.  The free-evolution period $T$ can, in principle, be extended far beyond the oscillator's decoherence limit, not only for differential frequency measurements~\cite{Chwalla2007Corr,Olmshenk2007Corr,CWC2011CorrSpec}, but also for clocks that keep time, as described here.  This raises the question of what the ultimate limits to atomic coherence are in clocks.  In early work, the millisecond transit time of particles in a thermal beam posed an artificial limit to coherence~\cite{Ramsey1956}.  For microwave frequencies, this was extended to about one second in atomic fountains~\cite{Kasevich1989}, and 100 seconds in ion traps~\cite{Berkeland1998}.  

Technical effects associated with fluctuating atomic frequencies pose one limit to atomic coherence at optical frequencies and above ($\omega>2\pi\times10^{14}$~Hz), but many of these can be reduced by experimental techniques.  Vacuum conditions pose another limit, because background-gas collisions can randomize the atomic phase, or eject atoms from the trap.  More fundamental is the excited-state lifetime, although some optical transitions have very long-lived excited states, such as the octupole transition in  Yb$^+$~\cite{Huntemann2012Octupole} and $^1$S$_0-^3$P$_0$ transitions in nuclear-spin zero isotopes~\cite{Taichenachev2006SpinZero}.
 
Each ensemble or sub-ensemble may be constructed in a separate chamber to avoid crosstalk, or a large collection of atoms can be subdivided into separate ensembles.  For the types of phase measurement considered here, the minimum number requirement for performance gain while maintaining six-sigma reliability is two ensembles of $36$ unentangled atoms or $19$ entangled atoms.  If the reduced-frequency clocks are generated by dynamical decoupling, the fidelity of $\pi$-pulses is a critical concern~\cite{Kirchmai2009HiFi}.  Preparation and read-out speed may limit the performance of reduced period clocks.  These difficulties are avoided if ensembles with different natural resonance frequencies can be combined. Technological barriers exist for all cases, but it is possible that future clock stability can surpass the limits of oscillator decoherence.

Helpful discussions with S.~Lloyd, D.~J.~Wineland, M.~D.~Lukin, E.~Knill, M.~Mullan, V.~Vuleti\'{c} and J.~A.~Sherman are gratefully acknowledged.  This work is supported by DARPA QuASAR, ONR, ARO, and AFOSR.  Not subject to U.S. copyright.

\bibliography{TR}
\end{document}